\documentclass[bj,authoryear,noshowframe]{imsart}

\usepackage{booktabs}
\usepackage{algorithm}
\usepackage{algorithmic}
\usepackage{placeins}
\startlocaldefs

\theoremstyle{plain}
\newtheorem{theorem}{Theorem}[section]
\newtheorem{lemma}{Lemma}[section]
\newtheorem{proposition}{Proposition}[section]
\newtheorem{corollary}{Corollary}[section]

\theoremstyle{definition}
\newtheorem{assumption}{Assumption}[section]

\newtheorem{example}{Example}[section]
\newtheorem{remark}{Remark}[section]

\endlocaldefs

\begin{document}

\begin{frontmatter}

\title{Sparse Spectral Precision Matrix Estimation for High-Dimensional Time Series}
\runtitle{Sparse spectral precision matrix estimation}

\begin{aug}

\author[A]{%
  \inits{L.}\fnms{Lili}~\snm{Wang}%
  \ead[label=e1]{liliwang@zjgsu.edu.cn}%
}
\author[A]{%
  \inits{L.}\fnms{Lang}~\snm{Shen}%
  \ead[label=e2]{24020040067@pop.zjgsu.edu.cn}%
}
\author[B]{%
  \inits{Z.}\fnms{Zhou}~\snm{Zhou}%
  \ead[label=e3]{zhou.zhou@utoronto.ca}%
}
\address[A]{%
  School of Statistics and Data Science,
  Zhejiang Gongshang University,
  Hangzhou 310018, China%
  \printead[presep={,\ }]{e1,e2}%
}

\address[B]{%
  Center for Data Science,
  Zhejiang University,
  Hangzhou, Zhejiang 310058, China%
  \printead[presep={,\ }]{e3}%
}

\end{aug}
\begin{abstract}
The spectral density matrix and its inverse, the spectral precision matrix,
characterize frequency-specific marginal and conditional linear dependence
structures, respectively. We develop a two-stage procedure for estimating
these matrices in high-dimensional settings. We estimate the spectral
density matrix with a kernel-smoothed periodogram estimator and then apply
the Spectral Graphical Lasso (SGLASSO) to estimate its sparse inverse.
SGLASSO uses a real-block representation to convert the complex Hermitian
inverse estimation problem into an \(\ell_1\)-penalized log-determinant program over
real symmetric positive definite matrices. Under a physical dependence
condition that allows the cumulative dependence measure to decay
polynomially without requiring Gaussianity or a linear process
representation, we establish a nonasymptotic elementwise maximum-norm error
bound for the kernel-smoothed periodogram estimator, uniformly over the
continuous frequency domain. We then derive uniform error bounds for SGLASSO
in the elementwise maximum, Frobenius, and spectral norms. When the nonzero
spectral precision entries are sufficiently separated from zero, we
establish simultaneous exact support recovery over any prescribed finite
set of frequencies with high probability. Numerical simulations show
competitive finite-sample performance in spectral precision matrix
estimation and graph recovery. An application to global bank volatility
data indicates geographically structured conditional dependence patterns
that vary across frequencies.
\end{abstract}


\end{frontmatter}

\section{Introduction}
The spectral density matrix plays a fundamental role in time series analysis,
characterizing the second-order properties of a multivariate stationary
process. A common approach for estimating this matrix is based on the
periodogram. However, the raw periodogram is generally inconsistent because
its variance does not vanish as the sample size increases, thus necessitating
smoothing across neighboring frequencies. In the classical regime where the
dimension is fixed, the asymptotic theory for smoothed periodogram estimators
is well established; see, for example, \cite{Brillinger1981,Brockwell1991,Hannan1970,Priestley1981}. 
With
the rapid advancement of data acquisition technologies, high-dimensional time
series have become increasingly common, which motivates the development of
new statistical methodology and theory to address the challenges of high
dimensionality. 
For example, several studies have applied shrinkage to
smoothed periodogram estimates; see, among others,
\citep{Boehm2009,Fiecas2011,Fiecas2014,SchneiderLuftman2016}. Under
sparsity assumptions, \citet{Sun2018} proposed thresholded versions of
averaged periodograms for estimating large spectral density matrices.
Theoretical results are also available when the dimension increases with the
sample size. \citet{WuZaffaroni2018} derived uniform convergence results for
spectral density estimates based on lag windows, while \citet{Fiecas2019}
established nonasymptotic error bounds for locally averaged periodogram estimators
under physical dependence.


However, these theoretical developments have largely focused on spectral
density matrices. The cross-spectral density between two component processes
describes their pairwise linear association without conditioning on the
remaining components. It therefore cannot determine whether an observed
association reflects a direct relationship or is mediated by other
components. Many applications instead concern the underlying conditional
dependence structure, which is characterized by the spectral precision
matrix, defined as the inverse of the spectral density matrix. Within the
graphical modeling framework introduced by \citet{Dahlhaus2000}, a
multivariate time series can be represented by a graph in which each vertex
corresponds to a component process. An edge is absent between two vertices
when the corresponding off-diagonal element of the spectral precision matrix
is zero over the entire frequency domain. At each frequency, the strength of
the conditional linear association can be quantified by partial coherence,
which is directly obtained from the spectral precision matrix. Partial
coherence and related frequency-domain graphical methods have been used to
study conditional dependence in financial networks
\citep{JimenezVaronKnight2026}, interactions among physiological variables in
intensive-care monitoring \citep{gather2002graphical}, and functional brain
connectivity in neuroscience
\citep{Fiecas2011,Medkour2009,salvador2005undirected}.

To estimate spectral precision matrices in high-dimensional settings,
regularization methods exploiting sparsity have been widely used. Using a
locally averaged periodogram at the Fourier frequencies, \citet{Fiecas2019} applied the constrained \(\ell_1\)-minimization approach of
\citet{Cai2011} to estimate the spectral precision matrix. More
recently, \citet{Deb2024} proposed a complex-valued extension of
graphical lasso methodology \citep{Yuan2007,Friedman2008}, based on a local
Whittle likelihood. However, Fiecas et al. assume geometric decay of the
physical dependence measure, whereas Deb et al. develop their theory for
Gaussian linear processes with short-range dependence. In contrast, we work
within the physical dependence framework introduced by \citet{Wu2005}
and allow the cumulative physical-dependence measure to decay polynomially,
thereby accommodating possibly nonlinear and non-Gaussian stationary
processes with more slowly decaying temporal dependence.

Our main contributions are summarized as follows. First, 
under the physical dependence framework, we establish a nonasymptotic elementwise maximum-norm error bound for the kernel-smoothed periodogram estimator, uniformly over the entire continuous frequency domain. Unlike the Fourier-grid result of \citet{Fiecas2019}, which was established under geometrically decaying physical dependence and an exponential moment condition, our result allows polynomial decay in the cumulative physical dependence measure.
Second, we propose SGLASSO for estimating sparse spectral precision matrices. Its real-block formulation permits the direct use of existing algorithms for solving the graphical lasso. Unlike the complex-modulus penalty employed by \citet{Deb2024}, SGLASSO applies an $\ell_1$ penalty componentwise to the real and imaginary parts of the off-diagonal entries, leading exactly to a classical
real-valued Graphical Lasso problem. We further prove that the resulting optimization problem admits a unique optimizer with the required real-block structure and that the complex reconstruction of this optimizer is Hermitian and positive definite. Third, we establish nonasymptotic error bounds for SGLASSO in the
elementwise maximum, Frobenius, and spectral norms, uniformly over
\([-\pi,\pi]\). In contrast to the fixed-frequency theory of
\citet{Deb2024} for Gaussian and general linear processes with short-range
dependence, our results accommodate possibly nonlinear and non-Gaussian
stationary processes under polynomially decaying cumulative physical
dependence.

The remainder of this paper is organized as follows. Section~2 introduces
the kernel-smoothed periodogram estimator of the spectral density matrix and
the SGLASSO estimator of the spectral precision matrix. Section~3 presents
uniform nonasymptotic error bounds for both estimators and a support recovery
result for SGLASSO. Section~4 reports numerical simulations evaluating
estimation accuracy and graph recovery.
Section~5 applies SGLASSO to global bank volatility data to investigate
frequency-specific conditional dependence networks. Proofs of all
theoretical results and the supporting auxiliary lemmas are provided in the
Supplementary Material \citep{Supplement2026}.

We use the following notation throughout the paper. Let
\(\mathbb Z\), \(\mathbb R\), and \(\mathbb C\) denote the sets of integers,
real numbers, and complex numbers, respectively. For \(x\in\mathbb R\),
\(\lfloor x\rfloor\) denotes the greatest integer not exceeding \(x\). For
\(a,b\in\mathbb R\), write \(a\vee b:=\max\{a,b\}\) and
\(a\wedge b:=\min\{a,b\}\). For an event \(E\), \(\mathbf 1\{E\}\) denotes
its indicator. The imaginary unit is denoted by
\(\mathrm i=\sqrt{-1}\). For \(z\in\mathbb C\), \(\Re(z)\), \(\Im(z)\),
and \(|z|\) denote its real part, imaginary part, and modulus, respectively.
For \(v=(v_1,\ldots,v_m)^\top\in\mathbb C^m\) and \(q\geq 1\), define
\(\|v\|_q:=(\sum_{i=1}^{m}|v_i|^q)^{1/q}\). For a scalar random variable
\(Y\) satisfying \(\mathbb E|Y|^q<\infty\), write
\(\|Y\|_q:=(\mathbb E|Y|^q)^{1/q}\). For
\(A=(A_{ij})\in\mathbb C^{m\times m}\), \(A^\top\) and \(A^*\) denote its
transpose and conjugate transpose, respectively, while
\(\operatorname{tr}(A)\) and \(\operatorname{vec}(A)\) denote its trace and
columnwise vectorization. We define
\(\|A\|_{\max}:=\max_{i,j}|A_{ij}|\),
\(\|A\|_F:=(\sum_{i,j}|A_{ij}|^2)^{1/2}\),
\(\|A\|_1:=\max_j\sum_i|A_{ij}|\), and
\(\|A\|_\infty:=\max_i\sum_j|A_{ij}|\), while \(\|A\|_2\) denotes the
spectral norm. The off-diagonal \(\ell_1\)-norm is defined by
\(\|A\|_{1,\mathrm{off}}:=\sum_{i\neq j}|A_{ij}|\), and the Kronecker
product is denoted by \(\otimes\). For any positive integer \(m\),
\(\mathbb S_{++}^m\) and \(\mathbb H_{++}^m\) denote the cones of
\(m\times m\) real symmetric positive definite matrices and
\(m\times m\) complex Hermitian positive definite matrices, respectively.

\section{Methodology}
\subsection{Kernel-Smoothed Periodogram Estimator}\label{subsec:KSP}
Let $\{X_t\}_{t\in\mathbb Z}$ be a $p$-dimensional, real-valued, zero-mean stationary process
with autocovariance matrix $\Gamma(\tau)=\mathbb E(X_{t+\tau}X_t^\top)$ satisfying
$\sum_{\tau\in\mathbb Z}\|\Gamma(\tau)\|_{\max}<\infty$. The spectral density matrix is defined by
\begin{equation}
f(\omega) = \frac{1}{2\pi}\sum_{\tau\in\mathbb Z}
\Gamma(\tau)\exp(-\mathrm i\tau\omega),
\qquad \omega\in[-\pi,\pi].
\end{equation}
Given observations $X_1,\ldots,X_n$, define the finite Fourier transform
\begin{equation}
d_n(\omega) = n^{-1/2}\sum_{t=1}^{n} X_t\exp(-\mathrm i t\omega),
\end{equation}
and the periodogram matrix
\begin{equation}
I_n(\omega) = \frac{1}{2\pi}\,d_n(\omega)d_n^*(\omega).
\end{equation}
For each fixed $\omega$, $I_n(\omega)$ is asymptotically unbiased for
$f(\omega)$ but is inconsistent, as its variance does not vanish as
$n\to\infty$
\citep{Brillinger1981,Priestley1981,Brockwell1991}. We therefore employ
the kernel-smoothed periodogram estimator
\begin{equation}\label{eq:kernel_cont}
    \tilde f(\omega) = \int_{-\pi}^{\pi} W_{h_n}(\omega-\theta)\,I_n(\theta)\,\mathrm{d}\theta,
\end{equation}
where
\begin{equation}\label{eq:weight}
    W_{h_n}(a)
    =
    \sum_{\ell\in\mathbb Z}
    \frac{1}{h_n}
    K\!\left(\frac{a+2\pi\ell}{h_n}\right).
\end{equation}
 Here \(K(\cdot)\) is a nonnegative, symmetric, and compactly supported
kernel satisfying \(\int_{\mathbb R}K(u)\,\mathrm du=1\), and \(h_n>0\)
is a bandwidth parameter. The full conditions on \(K(\cdot)\) and \(h_n\)
are specified in Section~\ref{sec:TR}. By construction, \(W_{h_n}(\cdot)\) is nonnegative and \(2\pi\)-periodic, and it
integrates to one over any interval of length \(2\pi\). 

\subsection{SGLASSO for Spectral Precision Matrices}\label{subsec:SGLASSO}
Suppose that \(f(\omega)\) is positive definite at every frequency under
consideration, and define the spectral precision matrix by
\(\Theta(\omega):=f(\omega)^{-1}\).
To estimate $\Theta(\omega)$, we use the kernel-smoothed periodogram estimator
$\tilde f(\omega)$ constructed in Section~\ref{subsec:KSP} as a plug-in
estimator of $f(\omega)$. In high-dimensional settings, $\tilde f(\omega)$ may
be singular or ill-conditioned, so its direct inverse may be undefined or
numerically unstable. We therefore estimate $\Theta(\omega)$ using an
$\ell_1$-penalized log-determinant formulation that exploits the assumed
sparsity of its off-diagonal entries.

The matrices $f(\omega)$, $\Theta(\omega)$, and $\tilde f(\omega)$ are
generally complex Hermitian, whereas the classical theory for the
$\ell_1$-penalized log-determinant estimator
\citep{Ravikumar2011} is formulated over the cone of real symmetric positive
definite matrices. To recast the regularized inverse estimation problem
within this framework, we introduce a real-block representation.
For any $Z\in\mathbb C^{p\times p}$, write $Z=A+\mathrm iB$, where
$A=\Re Z\in\mathbb R^{p\times p}$ and
$B=\Im Z\in\mathbb R^{p\times p}$, and define
\begin{equation}\label{eq:R_def}
    \mathcal R(Z)=
    \begin{pmatrix}
        A & -B\\[2pt]
        B & A
    \end{pmatrix}
    \in\mathbb R^{2p\times 2p}.
\end{equation}
The following lemma records the algebraic properties of the real block
representation used in the subsequent development. 

\begin{lemma}\label{lem:real-block}
The map \(\mathcal R(\cdot)\) has the following properties.
\begin{enumerate}
\item[(i)]
The map \(\mathcal R(\cdot)\) is real-linear and injective. For all
\(Z_1,Z_2\in\mathbb C^{p\times p}\),
\(\mathcal R(I_p)=I_{2p}\),
\(\mathcal R(Z_1Z_2)=\mathcal R(Z_1)\mathcal R(Z_2)\), and
\(\mathcal R(Z_1^*)=\mathcal R(Z_1)^\top\).
If \(Z\in\mathbb C^{p\times p}\) is invertible, then
\(\mathcal R(Z^{-1})=\mathcal R(Z)^{-1}\).

\item[(ii)]
For all \(Z_1,Z_2\in\mathbb C^{p\times p}\),
\[
    \operatorname{tr}\!\left(
        \mathcal R(Z_1)\mathcal R(Z_2)
    \right)
    =
    2\Re\!\left[
        \operatorname{tr}(Z_1Z_2)
    \right].
\]
For every \(Z\in\mathbb C^{p\times p}\),
\(\det(\mathcal R(Z))=|\det(Z)|^2\).
If \(Z\) is Hermitian positive definite, then
\(\log\det(\mathcal R(Z))=2\log\det(Z)\).

\item[(iii)]
A matrix \(Z\in\mathbb C^{p\times p}\) is Hermitian if and only if
\(\mathcal R(Z)\) is real symmetric. Moreover, \(Z\) is Hermitian
positive definite if and only if \(\mathcal R(Z)\) is real symmetric
positive definite.

\item[(iv)]
A matrix \(M\in\mathbb R^{2p\times2p}\) lies in the range of
\(\mathcal R(\cdot)\) if and only if \(J^\top M J=M\), or equivalently,
\(MJ=JM\), where
\[
    J:=
    \left(
    \begin{matrix}
        0&-I_p\\
        I_p&0
    \end{matrix}
    \right).
\]
In this case, the matrix \(Z\in\mathbb C^{p\times p}\) satisfying
\(M=\mathcal R(Z)\) is unique. If \(M\) is also symmetric, then the
corresponding \(Z\) is Hermitian.
\end{enumerate}
\end{lemma}
Applying the real-block representation to the spectral density and precision
matrices, we define
\begin{equation}\label{eq:ST_def}
\mathcal S(\omega):=\mathcal R(f(\omega)),\qquad
\mathcal T(\omega):=\mathcal R(\Theta(\omega)),\qquad
\tilde{\mathcal S}(\omega):=\mathcal R(\tilde f(\omega)).
\end{equation}
By Lemma~\ref{lem:real-block}, $\mathcal S(\omega)$ and
$\mathcal T(\omega)$ are real symmetric positive definite matrices, and
\(
\mathcal T(\omega)=\mathcal S(\omega)^{-1}
\).
Since $\tilde f(\omega)$ is Hermitian positive semidefinite,
$\tilde{\mathcal S}(\omega)$ is real symmetric positive semidefinite and
serves as the empirical input to the regularized inverse estimation problem.
At each frequency under consideration, we assume that
\(
\tilde f_{jj}(\omega)>0
\)
for all $j=1,\ldots,p$, or equivalently, that
$\tilde{\mathcal S}(\omega)$ has strictly positive diagonal entries.

For each \(\omega\in[-\pi,\pi]\), let
\(\lambda_{n,\omega}>0\) denote the tuning parameter at frequency
\(\omega\). Following the \(\ell_1\)-penalized log-determinant formulation
of \citet{Ravikumar2011}, we define
\begin{equation}\label{eq:SGLASSO_obj}
\tilde{\mathcal T}(\omega)
=
\operatorname*{arg\,min}_{\Psi\in\mathbb S_{++}^{2p}}
\left\{
\operatorname{tr}\!\left(
\tilde{\mathcal S}(\omega)\Psi
\right)
-\log\det(\Psi)
+\lambda_{n,\omega}\|\Psi\|_{1,\mathrm{off}}
\right\}.
\end{equation}
By Lemma~\ref{lem:real-block},
\(J^\top\tilde{\mathcal S}(\omega)J
=\tilde{\mathcal S}(\omega)\).
The next proposition establishes the existence, uniqueness, and
real-block structure of the optimizer.

\begin{proposition}\label{prop:structure-preservation}
Suppose that \(\tilde f_{jj}(\omega)>0\) for all
\(j=1,\ldots,p\). Then, for any \(\lambda_{n,\omega}>0\), the optimization
problem in \eqref{eq:SGLASSO_obj} admits a unique minimizer
\(\tilde{\mathcal T}(\omega)\). Moreover,
\[
J^\top \tilde{\mathcal T}(\omega)J
=
\tilde{\mathcal T}(\omega),
\]
and there exists a unique
\(\tilde\Theta(\omega)\in\mathbb H_{++}^{p}\) such that
\(
\tilde{\mathcal T}(\omega)
=
\mathcal R\bigl(\tilde\Theta(\omega)\bigr)
\).
\end{proposition}

The real-domain penalty has a direct interpretation in terms of the real and
imaginary parts of the complex precision matrix. For any
\(Q\in\mathbb H_{++}^{p}\), we have
\begin{equation}\label{eq:real-block-penalty}
\|\mathcal R(Q)\|_{1,\mathrm{off}}
=
2\sum_{i\neq j}
\left(
|\Re Q_{ij}|+|\Im Q_{ij}|
\right).
\end{equation}
Since both \(\tilde f(\omega)\) and \(Q\) are Hermitian,
\(\operatorname{tr}(\tilde f(\omega)Q)\) is real.
Lemma~\ref{lem:real-block} gives
\begin{align}
&\operatorname{tr}\!\left(
\mathcal R(\tilde f(\omega))\mathcal R(Q)
\right)
-\log\det\!\left(\mathcal R(Q)\right)
+\lambda_{n,\omega}\|\mathcal R(Q)\|_{1,\mathrm{off}}
\nonumber\\
&\qquad=
2\left[
\operatorname{tr}\!\left(\tilde f(\omega)Q\right)
-\log\det(Q)
+\lambda_{n,\omega}\sum_{i\neq j}
\left(
|\Re Q_{ij}|+|\Im Q_{ij}|
\right)
\right].
\label{eq:real-complex-objective}
\end{align}
Together with Proposition~\ref{prop:structure-preservation}, this gives the
equivalent characterization
\begin{equation}\label{eq:SGLASSO_complex}
\tilde\Theta(\omega)
=
\operatorname*{arg\,min}_{Q\in\mathbb H_{++}^{p}}
\left\{
\operatorname{tr}\!\left(\tilde f(\omega)Q\right)
-\log\det(Q)
+\lambda_{n,\omega}\sum_{i\neq j}
\left(
|\Re Q_{ij}|+|\Im Q_{ij}|
\right)
\right\}.
\end{equation}
The penalty acts componentwise on the real and imaginary parts of the
off-diagonal entries. A complex off-diagonal entry is zero precisely when
both components are zero. This componentwise penalty differs from a modulus-based penalty on the
complex off-diagonal entries and leads directly to the real-domain
graphical-lasso formulation in \eqref{eq:SGLASSO_obj}.

The main advantages are computational efficiency and methodological
convenience. The componentwise penalty leads exactly to the real-domain
graphical-lasso formulation, allowing classical graphical-lasso algorithms
and theoretical tools to be applied directly. In contrast, the modulus-based
penalty couples the real and imaginary parts and becomes a group-type penalty
after realification, requiring a specialized optimization procedure.
A computation-time comparison is reported in the Supplementary Material \citep{Supplement2026}.

For an explicit blockwise reconstruction, partition the real-domain solution
as
\[
\tilde{\mathcal T}(\omega)
=
\begin{pmatrix}
\tilde A_1(\omega)&\tilde B_1(\omega)\\
\tilde B_2(\omega)&\tilde A_2(\omega)
\end{pmatrix},
\]
where the four blocks are $p\times p$ real matrices.
Proposition~\ref{prop:structure-preservation} implies that the exact optimizer
satisfies
\(
\tilde A_1(\omega)=\tilde A_2(\omega)
\)
and
\(
\tilde B_2(\omega)=-\tilde B_1(\omega)
\).
Its complex counterpart can be written as
\begin{equation}\label{eq:Theta_recover}
\tilde\Theta(\omega)
=
\frac{1}{2}
\left(
\tilde A_1(\omega)+\tilde A_2(\omega)
\right)
+
\frac{\mathrm i}{2}
\left(
\tilde B_2(\omega)-\tilde B_1(\omega)
\right).
\end{equation}
The averaged form in \eqref{eq:Theta_recover} is also used in numerical
implementation to remove small finite-precision deviations from the exact
real-block structure. It satisfies
\[
\mathcal R\!\left(\tilde\Theta(\omega)\right)
=
\frac{1}{2}
\left(
\tilde{\mathcal T}(\omega)
+
J^\top\tilde{\mathcal T}(\omega)J
\right).
\]
If the real-domain matrix supplied to the reconstruction is symmetric positive
definite, then both matrices on the right-hand side are positive definite, and
so is their average. Lemma~\ref{lem:real-block}(iii) implies that
the reconstructed $\tilde\Theta(\omega)$ is Hermitian positive definite. For
the exact optimizer,
Proposition~\ref{prop:structure-preservation} reduces the preceding identity
to
\(
\mathcal R(\tilde\Theta(\omega))
=
\tilde{\mathcal T}(\omega)
\). 

Algorithm~\ref{alg:spec_glasso} summarizes the proposed estimation
procedure. The choices of \(h_n\) and \(\lambda_{n,\omega}\) used
in the numerical studies are described in
Section~\ref{sec:Simulation}.
\begin{algorithm}[htbp]
\caption{Spectral Precision Matrix Estimation via SGLASSO}
\label{alg:spec_glasso}
\begin{small}
\begin{algorithmic}
\STATE \textbf{Input:} centered observations
\(\{X_t\}_{t=1}^{n}\), kernel \(K(\cdot)\), bandwidth \(h_n\), a finite
target frequency set \(\Omega\subseteq[-\pi,\pi]\), and tuning parameters
\(\{\lambda_{n,\omega}>0:\omega\in\Omega\}\) 

\FOR{\(\omega\in\Omega\)}
    \STATE \(\tilde f(\omega)\leftarrow
    \int_{-\pi}^{\pi}
    W_{h_n}(\omega-\theta)I_n(\theta)\,\mathrm d\theta\)

    \STATE \textbf{Require:}
    \(\tilde f_{jj}(\omega)>0\) for all \(j=1,\ldots,p\)

    \STATE \(\tilde{\mathcal S}(\omega)
    \leftarrow\mathcal R(\tilde f(\omega))\)
    \hfill \(\triangleright\) real-block representation
    \eqref{eq:R_def}

    \STATE \(\tilde{\mathcal T}(\omega)\leftarrow\)
    the unique solution of \eqref{eq:SGLASSO_obj} with
    \(\lambda_{n,\omega}\)

    \STATE Partition
    \(
    \tilde{\mathcal T}(\omega)
    =
    \begin{pmatrix}
    \tilde A_1(\omega)&\tilde B_1(\omega)\\
    \tilde B_2(\omega)&\tilde A_2(\omega)
    \end{pmatrix}.
    \)

    \STATE \(\tilde\Theta(\omega)\leftarrow
    \frac{\tilde A_1(\omega)+\tilde A_2(\omega)}{2}
    +\mathrm{i}
    \frac{\tilde B_2(\omega)-\tilde B_1(\omega)}{2}\)
\ENDFOR

\STATE \textbf{return}
\(\{\tilde\Theta(\omega)\}_{\omega\in\Omega}\)
\end{algorithmic}
\end{small}
\end{algorithm}

\section{Theoretical Results}\label{sec:TR}
To study the theoretical properties of the proposed estimators, we consider a
general framework for high-dimensional stationary time series. The process
\(\{X_t\}_{t\in\mathbb Z}\) introduced in Section~\ref{subsec:KSP} is assumed to
admit the causal representation
\begin{equation}\label{eq:causal}
    X_t = G(\mathcal{H}_t)=(X_{t,1},\dots,X_{t,p})^\top,
\end{equation}
where \(\mathcal{H}_t=(\dots,\varepsilon_{t-1},\varepsilon_{t})\),
\(\{\varepsilon_t\}_{t\in\mathbb{Z}}\) are i.i.d.\ random elements, and
\(G(\cdot)=(g_1(\cdot),\dots,g_p(\cdot))^\top\) is an \(\mathbb{R}^p\)-valued
measurable function such that \(G(\mathcal{H}_t)\) is a well-defined random
vector. 

Following the physical dependence framework of \citet{Wu2005}, for
\(t\geq 0\) and \(1\leq j\leq p\), let
\(X_{t,j,\{0\}}:=g_j(\mathcal H_{t,\{0\}})\), where
\(\mathcal H_{t,\{0\}}\) is obtained from \(\mathcal H_t\) by replacing
\(\varepsilon_0\) with an independent copy
\(\tilde{\varepsilon}_0\). For \(q\geq2\), define the physical
dependence measure
\begin{equation}\label{eq:functionaldependence}
    \delta_{t,q,j}
    :=
    \bigl\|X_{t,j}-X_{t,j,\{0\}}\bigr\|_q,
\end{equation}
with the convention \(\delta_{t,q,j}=0\) for \(t<0\). This quantity
measures the effect of the innovation \(\varepsilon_0\) on the \(j\)-th
component at lag \(t\). Define the cumulative dependence measure
\begin{equation}\label{eq:cumulative_dependence}
    \Delta_{m,q,j}
    :=
    \sum_{t=m}^{\infty}\delta_{t,q,j},
    \qquad m\geq0.
\end{equation}
For \(q\geq 2\) and \(\beta>0\), we adopt the dependence-adjusted norm of
\citet{ZhangWu2021} defined by
\begin{equation}\label{eq:dependence_adjusted_q_norm}
    \|X_{\cdot,j}\|_{q,\beta}
    :=
    \sup_{m\geq 0}
    (m+1)^\beta\Delta_{m,q,j}.
\end{equation}
This norm incorporates the decay of the cumulative dependence measure into
the \(q\)-th moment control. The
parameter \(\beta\) determines the required polynomial decay rate:
a larger value of \(\beta\) requires faster decay with respect to the
lag \(m\).

\begin{assumption}\label{ass:DAEN}
There exist constants \(\vartheta\geq0\), \(\beta>0\), and
\(0<\mathcal K<\infty\), independent of \(n\) and \(p\), such that,
for every \(q\geq2\),
\begin{equation}\label{eq:DAEN_assumption}
    \max_{1\leq j\leq p}
    \|X_{\cdot,j}\|_{q,\beta}
    \leq
    \mathcal K q^\vartheta.
\end{equation}
\end{assumption}

Assumption~\ref{ass:DAEN} controls both temporal dependence and
moment growth. By \eqref{eq:dependence_adjusted_q_norm}, it implies
\(
    \Delta_{m,q,j}
    \leq
    \mathcal K q^\vartheta(m+1)^{-\beta}
\)
for every \(m\geq0\), \(q\geq2\), and \(1\leq j\leq p\).
Thus, \(\beta\) controls the polynomial decay imposed with respect
to the lag \(m\), whereas \(\vartheta\) controls the permitted growth
with respect to the moment order \(q\).
This assumption can be relaxed to a fixed \(q\)-th moment condition (\(q>4\)); see Proposition \ref{prop:fixed-q-nagaev}. 
The resulting bound contains an additional polynomial-tail term, including corresponding revisions to the convergence rates and subsequent SGLASSO theory.

\begin{assumption}\label{ass:bandwidth}
The kernel \(K(\cdot):\mathbb R\to\mathbb R\) is nonnegative,
symmetric, and bounded, with compact support \([-1,1]\) and
\(
    \int_{-1}^{1}K(u)\,\mathrm du=1
\).
The bandwidth \(h_n\) satisfies
\(
    0<h_n\leq H\) and \(nh_n\geq C_0
\),
for some fixed constants \(H,C_0>0\).
\end{assumption}

The kernel conditions are satisfied by many commonly used kernels,
including the Bartlett, uniform, and Epanechnikov kernels. The
restriction \(h_n\leq H\) keeps the smoothing bandwidth uniformly
bounded above, while \(nh_n\geq C_0\) rules out
\(h_n=o(n^{-1})\). The results are stated as nonasymptotic error bounds, for which \(h_n\to0\) is not required. This condition becomes necessary when the bounds are used to establish consistency.

\begin{theorem}\label{th:densitybound}
Suppose that Assumptions~\ref{ass:DAEN}
and~\ref{ass:bandwidth} hold. Then, for every \(\gamma>0\), there
exist constants \(C_1,C_2>0\), depending only on
\(\gamma\), \(\vartheta\), \(\beta\), \(K(\cdot)\), \(H\), and
\(C_0\), such that, with probability at least \(1-n^{-\gamma}\),
\begin{equation}\label{eq:density_bound}
    \sup_{\omega\in[-\pi,\pi]}
    \bigl\|
        \tilde f(\omega)-f(\omega)
    \bigr\|_{\max}
    \leq
    C_1\mathcal K^2 b_\beta(h_n)
    +
    C_2\mathcal K^2
    \frac{
        [1+\log(np^2)]^{2\vartheta+1}
    }{
        \sqrt{nh_n}
    },
\end{equation}
where
\begin{equation}\label{eq:b_beta}
    b_\beta(h_n)
    :=
    \begin{cases}
        h_n^\beta,
        & 0<\beta<2,\\[2mm]
        h_n^2\bigl[1+(\log(h_n^{-1})\vee0)\bigr],
        & \beta=2,\\[2mm]
        h_n^2,
        & \beta>2.
    \end{cases}
\end{equation}
\end{theorem}

\begin{remark}
Theorem~\ref{th:densitybound} provides a nonasymptotic bound for the
elementwise maximum-norm error of \(\tilde f(\omega)\), uniformly over
the entire continuous frequency domain \([-\pi,\pi]\). The first term
on the right-hand side of \eqref{eq:density_bound} is the bias
component determined jointly by temporal dependence and
frequency-domain smoothing, whereas the second term is the stochastic
component. The decay parameter \(\beta\) enters the bias component explicitly, whereas,
for each fixed \(\beta>0\), the stochastic component has the same dependence
on \(n\), \(p\), and \(h_n\). 

When \(\beta>2\), balancing the bias \(h_n^2\) with the stochastic
term in \eqref{eq:density_bound} gives the rate-balancing bandwidth
\begin{equation}\label{eq:bandwidth_beta_gt_2}
    h_n
    \asymp
    \left[
        \frac{\left(1+\log(np^2)\right)^{4\vartheta+2}}{n}
    \right]^{1/5}.
\end{equation}
Under this choice,
\begin{equation}\label{eq:density_rate_beta_gt_2}
    \sup_{\omega\in[-\pi,\pi]}
    \bigl\|
        \tilde f(\omega)-f(\omega)
    \bigr\|_{\max}
    =
    O_{\mathbb P}\!\left[
        \mathcal K^2
        \left(
            \frac{(1+\log(np^2))^{4\vartheta+2}}{n}
        \right)^{2/5}
    \right].
\end{equation}
Consequently, uniform max-norm consistency holds whenever
\(
    (1+\log(np^2))^{4\vartheta+2}=o(n)
\),
or equivalently,
\(
    \log p=o\!\left(n^{1/(4\vartheta+2)}\right).
\)
Under this condition, the bandwidth in
\eqref{eq:bandwidth_beta_gt_2} also satisfies \(h_n\to0\) and
\(nh_n\to\infty\). Apart from the logarithmic factor associated with
uniform high-probability control, the bandwidth and estimation error
have the orders \(n^{-1/5}\) and \(n^{-2/5}\), respectively. These
orders coincide with the classical pointwise bandwidth and
root-mean-square error orders for kernel-smoothed periodogram
estimators in the fixed-dimensional setting
\citep{Brillinger1981,Priestley1981,Brockwell1991}.

Closely related high-dimensional results were obtained by \citet{Fiecas2019}, as well as by \citet{ZhangZhang2025}. Under
geometrically decaying dependence, \citet{ZhangZhang2025} derived a uniform concentration bound for Welch's
non-overlapping segment-averaged spectral density estimator, whereas \citet{Fiecas2019} established a nonasymptotic error bound for a
locally averaged periodogram estimator on the Fourier-frequency grid. In
contrast, Theorem~\ref{th:densitybound} is established under the polynomial
decay condition in Assumption~\ref{ass:DAEN}, thereby allowing the cumulative
dependence measure to decay more slowly.
\end{remark}

Theorem~\ref{th:densitybound} requires the moment growth condition in
Assumption~\ref{ass:DAEN} for all \(q\ge2\).
The following proposition gives a uniform Nagaev-type bound for the
same estimator \(\tilde f\) under a dependence-adjusted moment condition
at a fixed order \(q>4\) and an additional decay condition on
the inverse Fourier transform of the kernel.

\begin{proposition}\label{prop:fixed-q-nagaev}
For some fixed \(q>4\) and \(\beta>0\), suppose that
\begin{equation}\label{eq:fixed-q-dependence}
    \max_{1\leq j\leq p}
    \|X_{\cdot,j}\|_{q,\beta}
    \leq
    \mathcal K_q,
\end{equation}
where \(0<\mathcal K_q<\infty\) is independent of \(n\) and \(p\).
Assume that \(K(\cdot)\) satisfies the kernel conditions in
Assumption~\ref{ass:bandwidth} and that, for fixed constants
\(C_K>0\) and \(\nu\geq1\), independent of \(n\) and \(p\),
\begin{equation}\label{eq:kernel-fourier-decay}
    \left|
        \int_{\mathbb R}K(v)e^{\mathrm iuv}\,\mathrm dv
    \right|
    \leq
    C_K(1+|u|)^{-\nu},
    \qquad u\in\mathbb R.
\end{equation}
Let \(h_n\) be deterministic and satisfy
\(0<h_n\leq1\) and \(nh_n\geq2\).
Then there exist constants \(C_3\geq1\), \(C_4>0\), and
\(c_1>0\), depending only on \(q\), \(\beta\), \(\nu\),
\(C_K\), and \(K(\cdot)\), such that, for every \(x>0\),
\begin{align}
&\mathbb P\!\left\{
    \sup_{\omega\in[-\pi,\pi]}
    \bigl\|
        \tilde f(\omega)-\mathbb E\tilde f(\omega)
    \bigr\|_{\max}
    >x
\right\}
\nonumber\\
&\quad\leq
    \frac{
        C_3p^2\mathcal K_q^q n^{1-q/2}
        \bigl[
            S_\nu(n,h_n)+D_\beta(n,h_n)
        \bigr]^{q/2}
    }{
        x^{q/2}
    }
    +
    \frac{C_3p^2}{h_n}
    \exp\!\left(
        -\frac{c_1nh_nx^2}{\mathcal K_q^4}
    \right),
    \label{eq:fixed-q-nagaev}
\end{align}
where
\begin{equation}\label{eq:fourier-shell-scale}
    S_\nu(n,h_n)
    :=
    \begin{cases}
        h_n^{-1},
        & \nu>1,\\[2mm]
        h_n^{-1}[1+\log(nh_n)]^{1+2/q},
        & \nu=1,
    \end{cases}
\end{equation}
and
\begin{equation}\label{eq:dependence-correction}
    D_\beta(n,h_n)
    :=
    h_n^{-(1/2+2/q)}
    \begin{cases}
        0,
        & \displaystyle \beta>1/2-2q^{-1},\\
        [1+\log(nh_n)]^{1+2/q},
        & \displaystyle \beta=1/2-2q^{-1},\\
        n^{1/2-2/q-\beta},
        & \displaystyle 0<\beta<1/2-2q^{-1}.
    \end{cases}
\end{equation}
Moreover, for every \(0<\eta<1\), with probability at least
\(1-\eta\),
\begin{align}
&\sup_{\omega\in[-\pi,\pi]}
    \bigl\|
        \tilde f(\omega)-f(\omega)
    \bigr\|_{\max}
\nonumber\\
&\quad\leq
    C_4\mathcal K_q^2
    \Bigg[
        b_\beta(h_n)
        +
        n^{2/q-1}
        \bigl[
            S_\nu(n,h_n)+D_\beta(n,h_n)
        \bigr]
        \left(\frac{p^2}{\eta}\right)^{2/q}
        +
        \sqrt{
            \frac{
                \log\{C_3p^2/(h_n\eta)\}
            }{
                nh_n
            }
        }
    \Bigg],
    \label{eq:fixed-q-high-probability}
\end{align}
where \(b_\beta(h_n)\) is defined in \eqref{eq:b_beta}.
\end{proposition}

\begin{remark}\label{rem:fixed-q-nagaev}
Proposition~\ref{prop:fixed-q-nagaev} provides uniform error bounds
for \(\tilde f\) under a dependence-adjusted moment condition at
a fixed order \(q>4\) and the polynomial decay condition
\eqref{eq:kernel-fourier-decay} on the Fourier transform of
\(K(\cdot)\).
The polynomial tail term in \eqref{eq:fixed-q-nagaev}
contributes the factor \((p^2/\eta)^{2/q}\) to the
high-probability error bound in
\eqref{eq:fixed-q-high-probability}.

For stationary processes, Proposition~4.3(i) of
\citet{ZhangWu2021} gives a uniform Nagaev-type bound for
a lag-window spectral density estimator based on sample
autocovariances at lags up to \(m\).
The present proposition applies directly to the continuous
estimator \(\tilde f(\omega)\) in \eqref{eq:kernel_cont},
whose lag weights are determined by the Fourier transform
of \(K(\cdot)\) and are retained at all sample lags.
Under \eqref{eq:fixed-q-dependence}, the polynomial term
in their bound is bounded by a constant multiple of
\(p^2\mathcal K_q^q n^{1-q/2}m^{q/2}x^{-q/2}\)
when \(\beta>1/2-2/q\).
For \(m\asymp h_n^{-1}\) satisfying their conditions on
the window width, this upper bound has the same order
as the polynomial term in \eqref{eq:fixed-q-nagaev}
when \(\nu>1\).
The present bound also covers the critical dependence case
\(\beta=1/2-2/q\) and the case \(\nu=1\);
the associated logarithmic factors are given in
\eqref{eq:dependence-correction} and
\eqref{eq:fourier-shell-scale}, respectively.
\end{remark}

Having established a uniform error bound for the kernel-smoothed
periodogram estimator in Theorem~\ref{th:densitybound}, we now study
the theoretical properties of the SGLASSO estimator
\(\tilde\Theta(\omega)\). Let
\(\mathcal V_p:=\{1,\ldots,p\}\). For each
\(\omega\in[-\pi,\pi]\), define the off-diagonal support of
\(\Theta(\omega)\) by
\begin{equation}\label{eq:support_def}
\mathcal E_\omega
:=
\left\{
(i,j)\in\mathcal V_p\times\mathcal V_p
\;\middle|\;
i\ne j,\ 
\Theta_{ij}(\omega)\ne0
\right\}.
\end{equation}
Let \(s_\omega:=|\mathcal E_\omega|\), where \((i,j)\) and
\((j,i)\) are counted separately. Define the maximum row support size by
\begin{equation}\label{eq:row_support_def}
d_\omega
:=
\max_{i\in\mathcal V_p}
\left|
\left\{
j\in\mathcal V_p
\;\middle|\;
\Theta_{ij}(\omega)\ne0
\right\}
\right|,
\end{equation}
where the diagonal entry is included. The corresponding uniform
sparsity quantities are
\begin{equation}\label{eq:global_sparsity}
s^{\max}
:=
\sup_{\omega\in[-\pi,\pi]}s_\omega,
\qquad
d^{\max}
:=
\sup_{\omega\in[-\pi,\pi]}d_\omega.
\end{equation}
Thus, \(s^{\max}\) controls the overall off-diagonal sparsity, whereas
\(d^{\max}\) controls the maximum row-wise sparsity, uniformly over
\(\omega\in[-\pi,\pi]\). For later use, define the estimated
off-diagonal support by
\begin{equation}\label{eq:estimated_support_def}
\tilde{\mathcal E}_\omega
:=
\left\{
(i,j)\in\mathcal V_p\times\mathcal V_p
\;\middle|\;
i\ne j,\ 
\tilde\Theta_{ij}(\omega)\ne0
\right\}.
\end{equation}

Because the SGLASSO optimization is carried out using the real block
representation, let
\(\mathcal V_{2p}:=\{1,\ldots,2p\}\) and define
\begin{equation}\label{eq:real_support_def}
\mathcal A_\omega
:=
\left\{
(i,j)\in\mathcal V_{2p}\times\mathcal V_{2p}
\;\middle|\;
\mathcal T_{ij}(\omega)\ne0
\right\},
\qquad
\mathcal A_\omega^c
:=
\left(
\mathcal V_{2p}\times\mathcal V_{2p}
\right)
\setminus\mathcal A_\omega.
\end{equation}
Since \(\mathcal T(\omega)\succ0\), every diagonal pair
\((i,i)\), \(i\in\mathcal V_{2p}\), belongs to
\(\mathcal A_\omega\). Hence, \(\mathcal A_\omega^c\) contains only
off-diagonal coordinate pairs corresponding to zero entries of
\(\mathcal T(\omega)\).
Recall that
\(\mathcal S(\omega)=\mathcal R(f(\omega))\) and
\(\mathcal T(\omega)=\mathcal S(\omega)^{-1}\).
The Hessian operator of
\(\Psi\mapsto-\log\det(\Psi)\), evaluated at
\(\Psi=\mathcal T(\omega)\), maps a perturbation \(V\) to
\(\mathcal S(\omega)V\mathcal S(\omega)\). Under columnwise
vectorization,
\[
\operatorname{vec}
\left(
\mathcal S(\omega)V\mathcal S(\omega)
\right)
=
\left(
\mathcal S(\omega)^\top\otimes\mathcal S(\omega)
\right)
\operatorname{vec}(V)
=
\left(
\mathcal S(\omega)\otimes\mathcal S(\omega)
\right)
\operatorname{vec}(V),
\]
where the last equality follows from the symmetry of
\(\mathcal S(\omega)\). Accordingly, define
\begin{equation}\label{eq:Hessian_def}
\Xi(\omega)
:=
\mathcal S(\omega)\otimes\mathcal S(\omega)
\in\mathbb R^{4p^2\times4p^2}.
\end{equation}
Under this vectorization convention, a coordinate pair
\((i,j)\in\mathcal V_{2p}\times\mathcal V_{2p}\) corresponds to
position \(i+2p(j-1)\). For
\(A,B\subseteq\mathcal V_{2p}\times\mathcal V_{2p}\), let
\(\Xi_{AB}(\omega)\) denote the
\(|A|\times|B|\) submatrix of \(\Xi(\omega)\) whose row indices
correspond to the pairs in \(A\) and whose column indices correspond
to the pairs in \(B\), with the ordering inherited from
\(\operatorname{vec}(\cdot)\).

Define
\begin{equation}\label{eq:phi_constants}
\phi_{\mathcal S}^{\max}
:=
\sup_{\omega\in[-\pi,\pi]}
\left(
1\vee\|\mathcal S(\omega)\|_\infty
\right),
\qquad
\phi_\Xi^{\max}
:=
\sup_{\omega\in[-\pi,\pi]}
\left(
1\vee
\left\|
\left(
\Xi_{\mathcal A_\omega\mathcal A_\omega}(\omega)
\right)^{-1}
\right\|_\infty
\right).
\end{equation}
These quantities are uniform over frequency but may depend on \(n\)
and \(p\). Beyond Assumptions~\ref{ass:DAEN}
and~\ref{ass:bandwidth}, the analysis of
\(\tilde\Theta(\omega)\) requires two additional conditions on the
population spectral density and precision matrices.
The quantity \(\phi_{\mathcal S}^{\max}\) controls, uniformly over
frequency, the maximum absolute row-sum norm of \(\mathcal S(\omega)\) and is used
to bound the higher-order remainder in the inverse expansion.
The quantity \(\phi_\Xi^{\max}\) controls the maximum absolute
row-sum norm of the inverse Hessian restricted to \(\mathcal A_\omega\) and measures how perturbations are
amplified in the restricted SGLASSO problem.
The truncation by one is a technical normalization that allows the
same scaling condition to control both the lower-order perturbation
terms and the higher-order remainder.

\begin{assumption}\label{ass:eigen}
There exist constants
\(0<\Lambda_{\min}\le\Lambda_{\max}<\infty\), independent of \(n\) and \(p\),
such that
\[
    \Lambda_{\min}
    \le
    \lambda_{\min}\bigl(f(\omega)\bigr)
    \le
    \lambda_{\max}\bigl(f(\omega)\bigr)
    \le
    \Lambda_{\max}
\]
for every \(\omega\in[-\pi,\pi]\).
\end{assumption}

Assumption~\ref{ass:eigen} requires the eigenvalues of \(f(\omega)\) to be
uniformly bounded away from zero and infinity over
\(\omega\in[-\pi,\pi]\). The lower bound ensures that
\(\Theta(\omega)=f(\omega)^{-1}\) is well defined, while the upper bound
controls the spectral norm of \(f(\omega)\). Together with
Lemma~\ref{lem:real-block}, this implies that
\(\mathcal S(\omega)\) and \(\mathcal T(\omega)=\mathcal S(\omega)^{-1}\)
are real symmetric positive definite matrices. Hence,
\(\Xi(\omega)\) is positive
definite, and its principal submatrix
\(\Xi_{\mathcal A_\omega\mathcal A_\omega}(\omega)\) is invertible for every
\(\omega\in[-\pi,\pi]\).

\begin{assumption}\label{ass:irrepresentability} There exists a constant \(\alpha\in(0,1]\), independent of \(n\) and \(p\), such that
\begin{equation}\label{eq:thm2-irrep}
    \sup_{\omega\in[-\pi,\pi]}
    \left\|
        \Xi_{\mathcal A_\omega^c\mathcal A_\omega}(\omega)
        \bigl(
            \Xi_{\mathcal A_\omega\mathcal A_\omega}(\omega)
        \bigr)^{-1}
    \right\|_\infty
    \leq
    1-\alpha.
\end{equation} 
\end{assumption}

Assumption~\ref{ass:irrepresentability} is the frequency-uniform analogue of
the irrepresentability condition used for the
\(\ell_1\)-penalized log-determinant estimator introduced by \citet{Ravikumar2011}. At each frequency \(\omega\),
\(\mathcal A_\omega\) indexes the nonzero entries of
\(\mathcal T(\omega)\), while \(\mathcal A_\omega^c\) indexes its zero
entries. The matrix
\(
  \Xi_{\mathcal A_\omega^c\mathcal A_\omega}(\omega)
    \bigl(
        \Xi_{\mathcal A_\omega\mathcal A_\omega}(\omega)
    \bigr)^{-1}
\)
measures the extent to which the coordinates outside
\(\mathcal A_\omega\) can be linearly represented by the coordinates in
\(\mathcal A_\omega\), with the representation determined by the Hessian.
The underlying requirement is that no zero coordinate of
\(\mathcal T(\omega)\) is too strongly associated, through the Hessian, with
the collection of its nonzero coordinates.

\begin{example}\label{ex:irrepresentability}
For every \(p\geq3\), consider the stationary 
VAR(1) process
\[
    X_t
    =
    aI_pX_{t-1}+\zeta_t,
    \qquad
    \operatorname{Cov}(\zeta_t)
    =
    C_p(\varrho),
    \qquad
    C_p(\varrho)
    :=
    \bigl(\varrho^{|j-k|}\bigr)_{j,k=1}^p,
\]
where \(I_p\) is the \(p\times p\) identity matrix and
\(\{\zeta_t\}_{t\in\mathbb Z}\) is a \(p\)-dimensional mean-zero
white noise process. Suppose that \(a,\varrho\in\mathbb R\),
\(|a|<1\), and \(0<\varrho<\sqrt{2}-1\), where \(a\) and
\(\varrho\) do not depend on \(n\) or \(p\).
The spectral density and spectral precision matrices are
\[
    f(\omega)
    =
    \frac{1}
    {2\pi|1-ae^{-\mathrm i\omega}|^2}
    C_p(\varrho),
    \qquad
    \Theta(\omega)
    =
    2\pi|1-ae^{-\mathrm i\omega}|^2
    C_p(\varrho)^{-1}.
\]
The inverse of \(C_p(\varrho)\) is
\[
    C_p(\varrho)^{-1}
    =
    \frac{1}{1-\varrho^2}
    \begin{pmatrix}
        1 & -\varrho & & & \\
        -\varrho & 1+\varrho^2 & -\varrho & & \\
        & \ddots & \ddots & \ddots & \\
        & & -\varrho & 1+\varrho^2 & -\varrho \\
        & & & -\varrho & 1
    \end{pmatrix}.
\]
Therefore, for every \(\omega\in[-\pi,\pi]\),
\[
    \mathcal E_\omega
    =
    \left\{
        (j,j+1),(j+1,j):
        j=1,\ldots,p-1
    \right\}.
\]
Let \(\overline C_p:=\operatorname{diag}
\{C_p(\varrho),C_p(\varrho)\}\).
The corresponding real-block matrices satisfy
\[
    \mathcal S(\omega)
    =
    \frac{1}
    {2\pi|1-ae^{-\mathrm i\omega}|^2}
    \overline C_p,
    \qquad
    \mathcal T(\omega)
    =
    2\pi|1-ae^{-\mathrm i\omega}|^2
    \overline C_p^{-1}.
\]
Thus, \(\mathcal A_\omega\) is independent of \(\omega\), we denote this
common support by \(\mathcal A\). Since
\[
    \Xi(\omega)
    =
    \bigl[
        2\pi|1-ae^{-\mathrm i\omega}|^2
    \bigr]^{-2}
    \bigl(
        \overline C_p\otimes\overline C_p
    \bigr),
\]
we have
\[
\begin{aligned}
    &\Xi_{\mathcal A^c\mathcal A}(\omega)
    \left[
        \Xi_{\mathcal A\mathcal A}(\omega)
    \right]^{-1}=(
        \overline C_p\otimes\overline C_p)_{\mathcal A^c\mathcal A}
    \left[(
            \overline C_p\otimes\overline C_p)_{\mathcal A\mathcal A}
    \right]^{-1}.
\end{aligned}
\]
Consider an inactive coordinate in either diagonal \(p\times p\)
block. If its two indices are \(i<j\), let \(d:=j-i\geq2\).
The corresponding row of
\(\Xi_{\mathcal A^c\mathcal A}(\omega)
[\Xi_{\mathcal A\mathcal A}(\omega)]^{-1}\)
has coefficient \(\varrho^{d-1}\) at each of the \(d\) edge
coordinates along the path from \(i\) to \(j\), coefficient
\(-\varrho^d\) at each of the \(d-1\) internal diagonal coordinates,
and zero at all other support coordinates. Its absolute row sum is
\(
    d\varrho^{d-1}
    +(d-1)\varrho^d
\).
The same expression applies when \(i>j\). The inactive coordinates
in the two off-diagonal \(p\times p\) blocks have zero rows in
\(\Xi_{\mathcal A^c\mathcal A}(\omega)
[\Xi_{\mathcal A\mathcal A}(\omega)]^{-1}\).

For every \(d\geq2\), the ratio of two consecutive absolute row sums
satisfies
\[
    \frac{
        (d+1)\varrho^d+d\varrho^{d+1}
    }{
        d\varrho^{d-1}+(d-1)\varrho^d
    }
    =
    \varrho
    \frac{
        d+1+d\varrho
    }{
        d+(d-1)\varrho
    }
    <
    2\varrho
    <
    1,
\]
where the first strict inequality follows from
\(d-1+(d-2)\varrho>0\). Hence, these absolute row sums decrease with
\(d\), and their maximum is attained at \(d=2\). Consequently,
\[
    \sup_{\omega\in[-\pi,\pi]}
    \left\|
        \Xi_{\mathcal A^c\mathcal A}(\omega)
        \left[
            \Xi_{\mathcal A\mathcal A}(\omega)
        \right]^{-1}
    \right\|_\infty
    =
    2\varrho+\varrho^2.
\]
Since \(0<\varrho<\sqrt{2}-1\),
Assumption~\ref{ass:irrepresentability} holds with
\(\alpha=1-2\varrho-\varrho^2>0\), independently of \(n\), \(p\),
and \(\omega\).
\end{example}


We now establish uniform nonasymptotic error bounds for the SGLASSO
estimator. Fix \(\gamma>0\), and let \(C_1\) and \(C_2\) denote the
corresponding constants in Theorem~\ref{th:densitybound}. For
notational convenience, define
\begin{equation}\label{eq:Delta_f}
\Delta_f
:=
C_1\mathcal K^2 b_\beta(h_n)
+
C_2\mathcal K^2
\frac{
[1+\log(np^2)]^{2\vartheta+1}
}{
\sqrt{nh_n}
}.
\end{equation}

\begin{theorem}\label{thm:consistency}
Suppose Assumptions~\ref{ass:DAEN}--\ref{ass:irrepresentability} hold.
Let \(c_\lambda\geq8/\alpha\) be a constant independent of \(n\) and
\(p\). Choose the tuning parameters
to satisfy
\begin{equation}\label{eq:lambda_range}
    \frac{8}{\alpha}\Delta_f
    \leq
    \lambda_{n,\omega}
    \leq
    c_\lambda\Delta_f
    \qquad
    \text{for every }\omega\in[-\pi,\pi],
\end{equation}
and suppose that
\begin{equation}\label{eq:scaling_condition}
    \Delta_f
    <
    \min\left\{
        \Lambda_{\min},
        \left[
            16\,d^{\max}
            (\phi_{\mathcal S}^{\max})^3
            (\phi_\Xi^{\max})^2
            C_\lambda^2
        \right]^{-1}
    \right\},
\end{equation}
where \(C_\lambda:=1+c_\lambda\).
Then, with probability at least \(1-n^{-\gamma}\),
\(\tilde\Theta(\omega)\) satisfies
\begin{align}
    \sup_{\omega\in[-\pi,\pi]}
    \bigl\|
        \tilde\Theta(\omega)-\Theta(\omega)
    \bigr\|_{\max}
    &\leq
    2\sqrt{2}\,
    C_\lambda\phi_\Xi^{\max}\Delta_f,
    \label{eq:bound_max}\\[2pt]
    \sup_{\omega\in[-\pi,\pi]}
    \bigl\|
        \tilde\Theta(\omega)-\Theta(\omega)
    \bigr\|_F
    &\leq
    2C_\lambda\phi_\Xi^{\max}
    \sqrt{p+2s^{\max}}\,
    \Delta_f,
    \label{eq:bound_F}\\[2pt]
    \sup_{\omega\in[-\pi,\pi]}
    \bigl\|
        \tilde\Theta(\omega)-\Theta(\omega)
    \bigr\|_2
    &\leq
    2\sqrt{2}\,
    C_\lambda\phi_\Xi^{\max}
    d^{\max}\Delta_f.
    \label{eq:bound_spectral}
\end{align}
\end{theorem}

\begin{remark}
Theorem~\ref{thm:consistency} transfers the uniform estimation error
of the kernel-smoothed periodogram estimator to the SGLASSO estimator.
The condition \(\Delta_f<\Lambda_{\min}\) guarantees that the SGLASSO
optimization problem admits a unique solution at every frequency.
Indeed, on the event in Theorem~\ref{th:densitybound},
\[
    \tilde f_{jj}(\omega)
    \geq
    f_{jj}(\omega)-\Delta_f
    \geq
    \Lambda_{\min}-\Delta_f
    >
    0
\]
for every \(1\leq j\leq p\) and
\(\omega\in[-\pi,\pi]\).
Proposition~\ref{prop:structure-preservation}
therefore ensures the existence and uniqueness of the SGLASSO solution.
The second restriction in \eqref{eq:scaling_condition}
keeps the inverse-expansion remainder sufficiently small and reflects
the joint effects of the estimation error, row-wise sparsity, and the
matrix quantities
\(\phi_{\mathcal S}^{\max}\) and \(\phi_\Xi^{\max}\).
If \(d^{\max}\), \(\phi_{\mathcal S}^{\max}\), or
\(\phi_\Xi^{\max}\) diverges, then
\(\Delta_f\to0\) alone is not sufficient for
\eqref{eq:scaling_condition}.
Since \(\Lambda_{\min}\) and \(C_\lambda\) are fixed positive constants,
and \(d^{\max}\), \(\phi_{\mathcal S}^{\max}\), and
\(\phi_\Xi^{\max}\) are at least one, a sufficient condition for
\eqref{eq:scaling_condition} to hold eventually is
\[
    d^{\max}
    (\phi_{\mathcal S}^{\max})^3
    (\phi_\Xi^{\max})^2
    \Delta_f
    \to0.
\]
\end{remark}

We next establish simultaneous exact support recovery over any prescribed
finite set of frequencies.

\begin{corollary}\label{cor:graph-recovery}
Suppose that the conditions of Theorem~\ref{thm:consistency} hold.
Let \(\Omega\subseteq[-\pi,\pi]\) be any prescribed nonempty finite set.
If
\begin{equation}\label{eq:min-signal}
    \min_{\omega\in\Omega}
    \min_{(i,j)\in\mathcal E_\omega}
    \left|\Theta_{ij}(\omega)\right|
    >
    2\sqrt{2}\,
    C_\lambda \phi_\Xi^{\max}\Delta_f,
\end{equation}
then, with probability at least \(1-n^{-\gamma}\),
\[
    \tilde{\mathcal E}_\omega
    =
    \mathcal E_\omega
    \qquad
    \text{for every }\omega\in\Omega.
\]
The minimum over an empty support is understood as \(+\infty\).
\end{corollary}
Condition~\eqref{eq:min-signal} is sufficient for support recovery, but we do not claim
that it is necessary. It ensures that every nonzero entry is
larger than the uniform estimation error. Moreover, since our results
hold uniformly over \([-\pi,\pi]\), \(\Omega\) may depend on \(n\) and
may be chosen as the full set of canonical Fourier frequencies, provided
that condition~\eqref{eq:min-signal} holds over this set.

\section{Numerical Simulation}\label{sec:Simulation}
In this section, we conduct Monte Carlo simulations to evaluate the
finite-sample performance of the proposed SGLASSO estimator. These experiments
complement the theoretical analysis by examining spectral precision matrix
estimation and graph recovery at the selected frequencies. We compare SGLASSO
with two benchmark estimators under four high-dimensional time series models
that differ in temporal dynamics, cross-sectional dependence structures, and
innovation mechanisms.

\subsection{Simulation Setup}

We consider sample sizes \(n\in\{200,400\}\) and dimensions
\(p\in\{50,100,200\}\) to evaluate the estimators under different levels of
dimensionality relative to the sample size.
For each combination of \(n\), \(p\), and the
data-generating process, we conduct 500 independent Monte Carlo
replications. Estimation is performed at the three target frequencies
\(\pi/4\), \(\pi/2\), and \(3\pi/4\). We compare the proposed SGLASSO
estimator with the CGLASSO-sc2 estimator of \citet{Deb2024} and
the inverse of the shrinkage spectral density matrix estimator developed by
\citet{Boehm2009}.

\subsubsection{Data-Generating Processes}
\noindent\textbf{(1) White noise.}
\begin{equation}
X_t=u_t,
\qquad
u_t\stackrel{\mathrm{i.i.d.}}{\sim}
\mathcal N_p(0,\Sigma_u).
\end{equation}
Let \(\Omega_u=\Sigma_u^{-1}\). Its off-diagonal entries satisfy
\((\Omega_u)_{i,i+1}=(\Omega_u)_{i+1,i}=0.30\) for
\(1\leq i\leq p-1\), and
\((\Omega_u)_{i,i+2}=(\Omega_u)_{i+2,i}=-0.15\) for
\(1\leq i\leq p-2\). All remaining off-diagonal entries are zero, and
\((\Omega_u)_{ii}=1+\sum_{j\neq i}|(\Omega_u)_{ij}|\).

\noindent\textbf{(2) VAR(1).}
\begin{equation}
X_t=AX_{t-1}+u_t,
\qquad
u_t\stackrel{\mathrm{i.i.d.}}{\sim}
\mathcal N_p(0,\Sigma_u^{(1)}).
\end{equation}
The matrix \(A\) is diagonal. For each \(k=0,\ldots,p/10-1\), its
diagonal entries satisfy \(A_{10k+r,10k+r}=0.25\) for
\(r=1,\ldots,5\) and \(A_{10k+r,10k+r}=-0.10\) for
\(r=6,\ldots,10\). Let
\(\Omega_u^{(1)}=(\Sigma_u^{(1)})^{-1}\). Its adjacent off-diagonal
entries satisfy
\((\Omega_u^{(1)})_{i,i+1}=(\Omega_u^{(1)})_{i+1,i}=0.308\) for
\(1\leq i\leq p-1\). In addition, for each
\(k=0,\ldots,p/10-1\) and \(r\in\{1,2,3\}\), set
\((\Omega_u^{(1)})_{10k+r,10k+r+5}
=(\Omega_u^{(1)})_{10k+r+5,10k+r}=0.176\).
All other off-diagonal entries are zero, and
\((\Omega_u^{(1)})_{ii}
=1+\sum_{j\neq i}|(\Omega_u^{(1)})_{ij}|\).

\noindent\textbf{(3) VARMA(2,2).}
\begin{equation}
X_t
=
A_1X_{t-1}
+
A_2X_{t-2}
+
u_t
+
B_1u_{t-1}
+
B_2u_{t-2},
\qquad
u_t\stackrel{\mathrm{i.i.d.}}{\sim}
\mathcal N_p(0,\Sigma_u^{(2)}).
\end{equation}
The coefficient matrices are
\(A_1=D_p(0.20,-0.08)\), \(A_2=D_p(0.03,0.02)\),
\(B_1=D_p(0.05,-0.025)\), and \(B_2=0.01I_p\), where
\begin{equation}
D_p(a,b)
=
I_{p/10}\otimes
\operatorname{diag}
(
\underbrace{a,\ldots,a}_{5},
\underbrace{b,\ldots,b}_{5}
).
\end{equation}
Let \(\Omega_u^{(2)}=(\Sigma_u^{(2)})^{-1}\). Its only nonzero
off-diagonal entries are adjacent entries, with
\((\Omega_u^{(2)})_{i,i+1}
=(\Omega_u^{(2)})_{i+1,i}=0.40\) when \(i\) is odd and \(0.22\)
when \(i\) is even, for \(1\leq i\leq p-1\). Its diagonal entries are
\((\Omega_u^{(2)})_{ii}
=1+\sum_{j\neq i}|(\Omega_u^{(2)})_{ij}|\).

\noindent\textbf{(4) Conditionally heteroskedastic VARMA(2,2)
(CH-VARMA(2,2)).}
\begin{equation}
\begin{aligned}
X_t
&=
A_1X_{t-1}
+
A_2X_{t-2}
+
\varepsilon_t
+
B_1\varepsilon_{t-1}
+
B_2\varepsilon_{t-2},\\
\varepsilon_t
&=
\sqrt{q_t}\,u_t,
\qquad
q_t
=
0.06
+
0.08\eta_{t-1}^{2}
+
0.86q_{t-1}.
\end{aligned}
\end{equation}
Here,
\(u_t\stackrel{\mathrm{i.i.d.}}{\sim}
\mathcal N_p(0,\Sigma_u^{(2)})\) and
\(\eta_t\stackrel{\mathrm{i.i.d.}}{\sim}\mathcal N(0,1)\), and the two
innovation sequences are mutually independent. The matrices
\(A_1\), \(A_2\), \(B_1\), \(B_2\), and \(\Sigma_u^{(2)}\) are the same
as those in the VARMA(2,2) setting.

\subsubsection{Estimation and Tuning}

The spectral density matrix is estimated using the kernel-smoothed
periodogram with the Bartlett kernel
\(K(u)=(1-|u|)\mathbf 1\{|u|\leq1\}\). We adopt the stabilized bandwidth
selection procedure of \citet{lee2001stabilized}, which selects a bandwidth
\(h_{n,j}\) separately for each component series \(j=1,\ldots,p\). The common
bandwidth \(h_n=p^{-1}\sum_{j=1}^p h_{n,j}\) is then used to estimate every
entry of the spectral density matrix. The same periodogram ordinates, kernel,
and bandwidth are used for all three methods.

At each target frequency, SGLASSO and CGLASSO-sc2 are fitted over their
respective regularization paths, each consisting of 30 logarithmically
spaced candidate values. Because both estimators are computed numerically,
an unordered off-diagonal pair \((i,j)\), with \(1\leq i<j\leq p\), is
treated as an estimated edge when the modulus of the corresponding estimated
spectral precision entry exceeds \(10^{-6}\). The BIC is computed using the
effective sample size induced by kernel smoothing, with the degrees of
freedom determined according to the real-valued parameterization of each
estimator. For each estimator and target frequency, the regularization
parameter \(\lambda_{n,\omega}\) is selected by minimizing the BIC among path
points whose estimated graphs contain between \(\lceil p/2\rceil\) and \(4p\)
edges. The same edge-count range is applied to both estimators throughout all
simulation settings and Monte Carlo replications. The shrinkage estimator
does not require regularization-parameter selection.

\subsubsection{Evaluation Metrics}

The finite-sample performance of the estimators is evaluated using the
relative Frobenius error (RFE), the F1 score, and the area under the receiver
operating characteristic curve (AUROC). RFE assesses spectral precision
matrix estimation, whereas the F1 score and AUROC assess graph recovery for
the BIC-selected estimate and over the full regularization path, respectively.
Because the shrinkage estimator does not produce a sparse regularization
path, RFE is reported for all three methods, while the F1 score and AUROC are
reported only for SGLASSO and CGLASSO-sc2.

At a target frequency \(\omega\), let \(\widehat\Theta(\omega)\) denote the
estimate produced by the method being evaluated. The relative Frobenius error
is defined as
\begin{equation}
\label{eq:simulation-rfe}
\operatorname{RFE}_{\omega}
=
\frac{
\|\widehat\Theta(\omega)-\Theta(\omega)\|_{\mathrm F}
}{
\|\Theta(\omega)\|_{\mathrm F}
}.
\end{equation}
For SGLASSO and CGLASSO-sc2, \(\widehat\Theta(\omega)\) is the BIC-selected
estimate; for the shrinkage method, it is the inverse of the shrinkage
spectral density estimate.

For graph-recovery evaluation, define the numerical true edge set at
frequency \(\omega\) by
\[
\mathcal G_\omega
:=
\left\{
(i,j)\in\mathcal V_p\times\mathcal V_p
\;\middle|\;
i<j,\ 
|\Theta_{ij}(\omega)|>10^{-8}
\right\},
\]
and let
\(
\mathcal G_\omega^c
:=
\left\{
(i,j)\in\mathcal V_p\times\mathcal V_p
\;\middle|\;
i<j
\right\}
\setminus\mathcal G_\omega
\).
The threshold \(10^{-8}\) removes floating-point roundoff arising from
the analytic evaluation of \(\Theta(\omega)\); entries with modulus no
greater than this threshold are treated as numerical zeros.

For each of SGLASSO and CGLASSO-sc2, let
\(\widehat{\mathcal G}_{\mathrm{BIC},\omega}\) denote the unordered edge set
obtained from the BIC-selected estimate. The following metrics are computed
separately for the two methods, with the method index suppressed. The numbers
of true positives, false positives, and false negatives are defined as
\begin{equation}
\label{eq:simulation-edge-counts}
\begin{aligned}
\operatorname{TP}_{\omega}
=
\left|
\widehat{\mathcal G}_{\mathrm{BIC},\omega}
\cap\mathcal G_\omega
\right|,\qquad
\operatorname{FP}_{\omega}
=
\left|
\widehat{\mathcal G}_{\mathrm{BIC},\omega}
\cap\mathcal G_\omega^c
\right|,\qquad
\operatorname{FN}_{\omega}
=
\left|
\mathcal G_\omega
\setminus
\widehat{\mathcal G}_{\mathrm{BIC},\omega}
\right|.
\end{aligned}
\end{equation}
The F1 score is defined as
\begin{equation}
\label{eq:simulation-f1}
\operatorname{F1}_{\omega}
=
\frac{
2\operatorname{TP}_{\omega}
}{
2\operatorname{TP}_{\omega}
+\operatorname{FP}_{\omega}
+\operatorname{FN}_{\omega}
}.
\end{equation}
The F1 score is set to zero when the denominator in
\eqref{eq:simulation-f1} is zero.

For AUROC, write
\(\Lambda_{n,\omega}
=\{\lambda_{n,\omega}^{(1)},\ldots,\lambda_{n,\omega}^{(30)}\}\)
for the candidate regularization grid of either SGLASSO or CGLASSO-sc2.
Let \(\widehat{\mathcal G}^{(\ell)}_\omega\) denote the estimated unordered
edge set obtained at \(\lambda_{n,\omega}^{(\ell)}\), using the numerical
edge rule specified above. The corresponding true-positive and false-positive
rates are
\begin{equation}
\label{eq:simulation-roc-rates}
\begin{aligned}
\operatorname{TPR}_{\ell,\omega}
=
\frac{
\left|
\widehat{\mathcal G}^{(\ell)}_\omega
\cap\mathcal G_\omega
\right|
}{
|\mathcal G_\omega|
},\qquad
\operatorname{FPR}_{\ell,\omega}
=
\frac{
\left|
\widehat{\mathcal G}^{(\ell)}_\omega
\cap\mathcal G_\omega^c
\right|
}{
|\mathcal G_\omega^c|
}.
\end{aligned}
\end{equation}
The ROC curve is constructed from the points
\((\operatorname{FPR}_{\ell,\omega},
\operatorname{TPR}_{\ell,\omega})\),
\(\ell=1,\ldots,30\), together with the endpoints
\((0,0)\) and \((1,1)\). AUROC is computed as the area under this curve.
Smaller RFE values and larger F1 and AUROC values indicate better performance.

\subsection{Simulation Results}
Tables~\ref{tab:sim-rfe}--\ref{tab:sim-auroc} summarize the RFE, F1, and
AUROC results, respectively. SGLASSO attained the lowest mean RFE in all
18 settings under the three dynamic models; under white noise, the shrinkage
estimator had lower RFE at \(n=200\), whereas SGLASSO had lower RFE at
\(n=400\). For the F1 score, SGLASSO performed better in all six white-noise
settings and five of the six VAR(1) settings, while CGLASSO-sc2 produced
modestly higher values under VARMA(2,2) and CH-VARMA(2,2). SGLASSO yielded
higher AUROC values in all 24 settings, indicating better overall
discrimination between true edges and nonedges along the regularization
path. Across the four designs, increasing \(n\) generally improved all three
metrics, while increasing \(p\) tended to increase RFE and reduce the F1
score. Overall, the results favor SGLASSO in spectral precision matrix
estimation and path-level edge discrimination, while the relative
performance of the BIC-selected support depends on the data-generating
process. Additional comparisons of computation times for SGLASSO and
CGLASSO-sc2 are reported in Section~S6 of the Supplementary Material
\citep{Supplement2026}.

\begin{table}[htbp]
\centering
\caption{RFE of the BIC-selected SGLASSO and CGLASSO-sc2 estimators and
the inverse of the shrinkage spectral density estimator. The results are
averaged over the three target frequencies. Monte Carlo standard
deviations are reported in parentheses. For each data-generating process
and each \((n,p)\) combination, the smallest mean RFE among the three
methods is shown in bold.}
{\scriptsize
\renewcommand{\arraystretch}{1.10}
\setlength{\tabcolsep}{2.4pt}
\begin{tabular}{ll ccc @{\hspace{6pt}} ccc}
\toprule
&& \multicolumn{3}{c}{$n=200$}
& \multicolumn{3}{c}{$n=400$}\\
\cmidrule(lr){3-5}\cmidrule(lr){6-8}
Setting & Method
& $p=50$ & $p=100$ & $p=200$
& $p=50$ & $p=100$ & $p=200$\\
\midrule
White noise
& SGLASSO
& 0.246\,(0.006)
& 0.250\,(0.004)
& 0.253\,(0.003)
& \textbf{0.205\,(0.010)}
& \textbf{0.219\,(0.007)}
& \textbf{0.230\,(0.005)}\\

& CGLASSO-sc2
& 0.252\,(0.005)
& 0.254\,(0.003)
& 0.257\,(0.002)
& 0.239\,(0.007)
& 0.242\,(0.003)
& 0.244\,(0.002)\\

& Shrinkage
& \textbf{0.228\,(0.004)}
& \textbf{0.239\,(0.002)}
& \textbf{0.246\,(0.002)}
& 0.211\,(0.003)
& 0.228\,(0.002)
& 0.239\,(0.001)\\
\addlinespace[1pt]

VAR(1)
& SGLASSO
& \textbf{0.289\,(0.006)}
& \textbf{0.293\,(0.004)}
& \textbf{0.295\,(0.003)}
& \textbf{0.261\,(0.007)}
& \textbf{0.269\,(0.004)}
& \textbf{0.274\,(0.003)}\\

& CGLASSO-sc2
& 0.291\,(0.006)
& 0.294\,(0.004)
& 0.297\,(0.003)
& 0.272\,(0.005)
& 0.275\,(0.003)
& 0.277\,(0.002)\\

& Shrinkage
& 0.313\,(0.004)
& 0.326\,(0.003)
& 0.333\,(0.002)
& 0.292\,(0.004)
& 0.312\,(0.002)
& 0.325\,(0.002)\\
\addlinespace[1pt]

VARMA(2,2)
& SGLASSO
& \textbf{0.296\,(0.007)}
& \textbf{0.302\,(0.005)}
& \textbf{0.306\,(0.003)}
& \textbf{0.257\,(0.008)}
& \textbf{0.269\,(0.005)}
& \textbf{0.277\,(0.004)}\\

& CGLASSO-sc2
& 0.301\,(0.006)
& 0.304\,(0.004)
& 0.308\,(0.003)
& 0.270\,(0.007)
& 0.277\,(0.004)
& 0.282\,(0.003)\\

& Shrinkage
& 0.328\,(0.005)
& 0.341\,(0.003)
& 0.349\,(0.003)
& 0.304\,(0.004)
& 0.326\,(0.003)
& 0.340\,(0.002)\\
\addlinespace[1pt]

CH-VARMA(2,2)
& SGLASSO
& \textbf{0.300\,(0.020)}
& \textbf{0.305\,(0.017)}
& \textbf{0.310\,(0.018)}
& \textbf{0.259\,(0.017)}
& \textbf{0.273\,(0.016)}
& \textbf{0.280\,(0.015)}\\

& CGLASSO-sc2
& 0.304\,(0.020)
& 0.307\,(0.017)
& 0.312\,(0.018)
& 0.273\,(0.016)
& 0.281\,(0.015)
& 0.284\,(0.014)\\

& Shrinkage
& 0.328\,(0.018)
& 0.339\,(0.016)
& 0.349\,(0.018)
& 0.304\,(0.013)
& 0.326\,(0.012)
& 0.338\,(0.012)\\
\bottomrule
\end{tabular}
}
\label{tab:sim-rfe}
\end{table}

\begin{table}[htbp]
\centering
\caption{F1 scores of the BIC-selected SGLASSO and CGLASSO-sc2
estimators. The results are averaged over the three target frequencies.
Monte Carlo standard deviations are reported in parentheses. For each
data-generating process and each \((n,p)\) combination, the higher mean
F1 score of the two methods is shown in bold.}
{\scriptsize
\renewcommand{\arraystretch}{1.10}
\setlength{\tabcolsep}{2.4pt}
\begin{tabular}{ll ccc @{\hspace{6pt}} ccc}
\toprule
&& \multicolumn{3}{c}{$n=200$}
& \multicolumn{3}{c}{$n=400$}\\
\cmidrule(lr){3-5}\cmidrule(lr){6-8}
Setting & Method
& $p=50$ & $p=100$ & $p=200$
& $p=50$ & $p=100$ & $p=200$\\
\midrule
White noise
& SGLASSO
& \textbf{0.413\,(0.044)}
& \textbf{0.365\,(0.030)}
& \textbf{0.315\,(0.024)}
& \textbf{0.651\,(0.043)}
& \textbf{0.587\,(0.041)}
& \textbf{0.521\,(0.030)}\\

& CGLASSO-sc2
& 0.373\,(0.030)
& 0.337\,(0.024)
& 0.292\,(0.020)
& 0.467\,(0.041)
& 0.438\,(0.022)
& 0.420\,(0.016)\\
\addlinespace[1pt]

VAR(1)
& SGLASSO
& \textbf{0.404\,(0.040)}
& \textbf{0.340\,(0.031)}
& 0.272\,(0.020)
& \textbf{0.588\,(0.041)}
& \textbf{0.522\,(0.029)}
& \textbf{0.462\,(0.023)}\\

& CGLASSO-sc2
& 0.397\,(0.038)
& 0.336\,(0.030)
& \textbf{0.275\,(0.020)}
& 0.545\,(0.031)
& 0.503\,(0.022)
& 0.460\,(0.019)\\
\addlinespace[1pt]

VARMA(2,2)
& SGLASSO
& 0.499\,(0.042)
& 0.427\,(0.031)
& 0.359\,(0.022)
& 0.655\,(0.034)
& 0.607\,(0.024)
& 0.551\,(0.018)\\

& CGLASSO-sc2
& \textbf{0.506\,(0.042)}
& \textbf{0.442\,(0.032)}
& \textbf{0.380\,(0.024)}
& \textbf{0.665\,(0.036)}
& \textbf{0.621\,(0.024)}
& \textbf{0.578\,(0.018)}\\
\addlinespace[1pt]

CH-VARMA(2,2)
& SGLASSO
& 0.482\,(0.042)
& 0.410\,(0.031)
& 0.342\,(0.024)
& 0.645\,(0.032)
& 0.590\,(0.026)
& 0.535\,(0.019)\\

& CGLASSO-sc2
& \textbf{0.491\,(0.043)}
& \textbf{0.425\,(0.033)}
& \textbf{0.362\,(0.025)}
& \textbf{0.655\,(0.035)}
& \textbf{0.609\,(0.027)}
& \textbf{0.562\,(0.018)}\\
\bottomrule
\end{tabular}
}
\label{tab:sim-f1}
\end{table}

\begin{table}[!htbp]
\centering
\caption{AUROC for SGLASSO and CGLASSO-sc2 along their respective
regularization paths. The results are averaged over the three target
frequencies. Monte Carlo standard deviations are reported in
parentheses. For each data-generating process and each \((n,p)\)
combination, the higher mean AUROC of the two methods is shown in bold.}
{\scriptsize
\renewcommand{\arraystretch}{1.10}
\setlength{\tabcolsep}{2.4pt}
\begin{tabular}{ll ccc @{\hspace{6pt}} ccc}
\toprule
&& \multicolumn{3}{c}{$n=200$}
& \multicolumn{3}{c}{$n=400$}\\
\cmidrule(lr){3-5}\cmidrule(lr){6-8}
Setting & Method
& $p=50$ & $p=100$ & $p=200$
& $p=50$ & $p=100$ & $p=200$\\
\midrule
White noise
& SGLASSO
& \textbf{0.817\,(0.023)}
& \textbf{0.817\,(0.017)}
& \textbf{0.814\,(0.012)}
& \textbf{0.900\,(0.017)}
& \textbf{0.898\,(0.013)}
& \textbf{0.898\,(0.008)}\\

& CGLASSO-sc2
& 0.809\,(0.022)
& 0.803\,(0.016)
& 0.795\,(0.012)
& 0.891\,(0.017)
& 0.888\,(0.012)
& 0.885\,(0.008)\\
\addlinespace[1pt]

VAR(1)
& SGLASSO
& \textbf{0.843\,(0.023)}
& \textbf{0.842\,(0.017)}
& \textbf{0.842\,(0.012)}
& \textbf{0.922\,(0.015)}
& \textbf{0.923\,(0.010)}
& \textbf{0.922\,(0.008)}\\

& CGLASSO-sc2
& 0.834\,(0.023)
& 0.830\,(0.017)
& 0.822\,(0.012)
& 0.909\,(0.016)
& 0.908\,(0.012)
& 0.906\,(0.009)\\
\addlinespace[1pt]

VARMA(2,2)
& SGLASSO
& \textbf{0.882\,(0.022)}
& \textbf{0.881\,(0.016)}
& \textbf{0.881\,(0.012)}
& \textbf{0.944\,(0.014)}
& \textbf{0.945\,(0.010)}
& \textbf{0.944\,(0.007)}\\

& CGLASSO-sc2
& 0.877\,(0.023)
& 0.872\,(0.017)
& 0.867\,(0.011)
& 0.934\,(0.017)
& 0.934\,(0.012)
& 0.931\,(0.008)\\
\addlinespace[1pt]

CH-VARMA(2,2)
& SGLASSO
& \textbf{0.878\,(0.023)}
& \textbf{0.876\,(0.016)}
& \textbf{0.876\,(0.012)}
& \textbf{0.941\,(0.015)}
& \textbf{0.940\,(0.010)}
& \textbf{0.941\,(0.007)}\\

& CGLASSO-sc2
& 0.872\,(0.023)
& 0.866\,(0.017)
& 0.860\,(0.013)
& 0.930\,(0.018)
& 0.929\,(0.012)
& 0.928\,(0.009)\\
\bottomrule
\end{tabular}
}
\label{tab:sim-auroc}
\end{table}

\begin{figure}[htbp]
    \setlength{\abovecaptionskip}{4pt}
    \centering
    \includegraphics[width=0.85\textwidth]{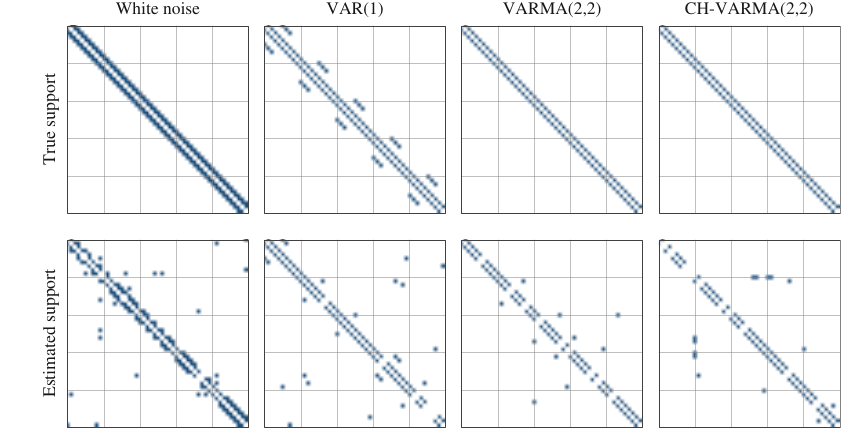}
    \caption{True and estimated off-diagonal supports at
    \(n=400\), \(p=50\), and \(\omega=\pi/2\).
    The upper row displays the true supports, and the lower row displays
    the supports estimated by the BIC-selected SGLASSO estimator.
    For illustration, the lower row displays the estimated support from the
same prespecified Monte Carlo replication for all four data-generating
processes. Diagonal entries are omitted.}
    \label{fig:support-recovery}
\end{figure}

Figure~\ref{fig:support-recovery} complements the aggregate results by
comparing the numerical true supports with the BIC-selected SGLASSO supports
at \(n=400\), \(p=50\), and \(\omega=\pi/2\). Across the four designs, the
estimated supports retain the dominant diagonal-banded patterns; under
VAR(1), they also recover several block-specific off-band connections,
although missed edges and isolated false positives remain. Because the lower
row uses the same fixed Monte Carlo replication index for all four designs,
the figure provides a visual illustration of one finite-sample realization,
while the overall graph-recovery performance is summarized by the F1 and
AUROC results reported above.

\section{Real Data Analysis}
To illustrate the proposed method, we consider the global bank volatility
data analyzed by \citet{Demirer2018}, which contains daily
range-based volatility measures for 96 major banks across the Americas,
Europe, Asia, Australia, and Africa. The original sample consists of 2676
trading days from 12 September 2003 to 7 February 2014. We use these data to
examine frequency-specific conditional linear associations among bank
volatility series.

After taking logarithms of the volatility measures, we use the final dyadic
block of \(n=2048\) observations, covering 17 February 2006 to 7 February
2014, for both stationarity screening and spectral estimation. We assess the
second-order stationarity of each series using the locally stationary wavelet
test developed by \citet{Nason2013}. The test applies a
false-discovery-rate adjustment to its coefficient-level hypotheses, and
we retain a bank when none is rejected at the 5\% level. This screening
yields 55 banks: 20 from Europe, 15 from the Americas, 14 from Asia,
5 from Australia, and 1 from Africa. Each retained log-volatility series
is centered and standardized to unit variance before spectral estimation.
For the resulting \(p=55\) panel, we apply the proposed method to estimate
the spectral precision matrix \(\Theta(\omega)\) and quantify pairwise
frequency-specific conditional linear associations using partial coherence
\citep{Medkour2009,Fiecas2011,SchneiderLuftman2016}. For
\(1\le i<j\le p\), the partial coherence between the \(i\)-th and \(j\)-th
component series at frequency \(\omega\) is defined as
\begin{equation}\label{eq:partial-coherence}
    \rho_{ij}(\omega)
    =
    \frac{
        |\Theta_{ij}(\omega)|^2
    }{
        \Theta_{ii}(\omega)\Theta_{jj}(\omega)
    }.
\end{equation}
The quantity \(\rho_{ij}(\omega)\in[0,1]\) measures the strength of their
linear association at frequency \(\omega\) after removing the linear
effects of all remaining series. 
The analysis focuses on three frequencies,
\(\omega=205\pi/512\), \(49\pi/512\), and \(33\pi/1024\),
corresponding approximately to cycles of one week, one month, and one
quarter, respectively. At these frequencies, we use the same
kernel-smoothed spectral density estimator and SGLASSO formulation as in
the simulation study, with the Bartlett kernel and data-driven choices of
\(h_n\) and \(\lambda_{n,\omega}\).

\begin{figure}[!htbp]
    \setlength{\abovecaptionskip}{4pt}
    \centering
    \includegraphics[width=\textwidth]{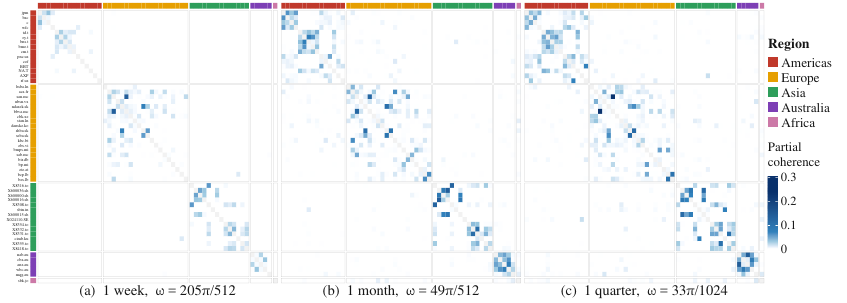}
    \caption{
    Estimated partial coherence matrices for the global bank volatility
    series at \(\omega=205\pi/512\), \(49\pi/512\), and \(33\pi/1024\),
    corresponding approximately to cycles of one week, one month, and one
    quarter, respectively. Banks are ordered and grouped by geographical
    region.A common color scale is used across the three panels, with darker shades indicating larger estimated partial coherence values. Diagonal entries are omitted.
    }
    \label{fig:bank-pc-heatmaps}
\end{figure}

Figure~\ref{fig:bank-pc-heatmaps} presents the estimated partial coherence
matrices, with the banks grouped by geographical region. The selected
graphs contain 123, 339, and 396 of the 1485 possible edges at the weekly,
monthly, and quarterly frequencies, respectively. The corresponding
ratios of within-region to cross-region edge density are \(53.6\),
\(4.0\), and \(3.5\). Thus, geographical concentration in the selected
support is most pronounced at the weekly frequency, whereas cross-region
conditional associations become more prevalent at the monthly and
quarterly frequencies.

\begin{figure}[!htbp]
    \setlength{\abovecaptionskip}{4pt}
    \centering
    \includegraphics[width=\textwidth]{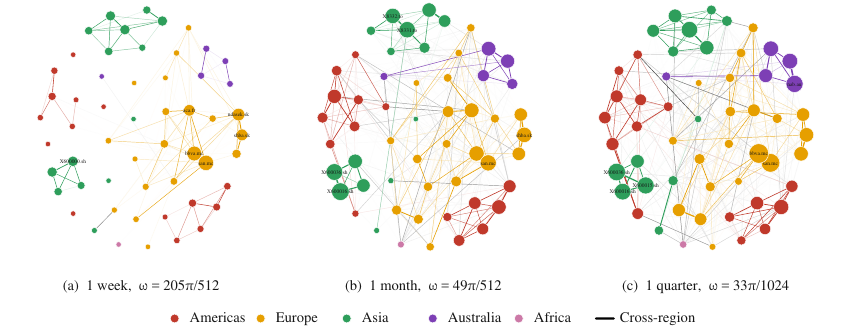}
    \caption{Estimated partial coherence networks at
    (a) \(\omega=205\pi/512\),
    (b) \(\omega=49\pi/512\), and
    (c) \(\omega=33\pi/1024\).
     Nodes are colored by geographical region, and node size is
    proportional to node strength. Edge width reflects the magnitude of
    partial coherence; black edges connect banks from different regions.
    The six banks with the largest node strengths are labeled in each
    panel. A common layout is used across the three panels, and node
    positions do not represent geographical locations.}
    \label{fig:bank-pc-networks}
\end{figure}

Figure~\ref{fig:bank-pc-networks} provides a network representation of
the estimated partial coherence matrices. Each node represents a bank,
an edge represents a selected nonzero partial coherence relationship,
and node size is proportional to node strength, defined as the sum of
the partial coherence values over all incident edges. Under the main
specification, Banco Santander (\texttt{san.mc}) is among the six nodes
with the largest strengths at all three frequencies. Banco Bilbao
Vizcaya Argentaria (\texttt{bbva.mc}), Svenska Handelsbanken
(\texttt{shba.sk}), China Merchants Bank (\texttt{X600036.sh}), and
China Minsheng Banking Corp (\texttt{X600016.sh}) appear among the six
largest-strength nodes at two frequencies. Their repeated appearance
indicates relatively broad conditional associations in the fitted
networks at more than one horizon.
\FloatBarrier

Taken together, Figures~\ref{fig:bank-pc-heatmaps} and
\ref{fig:bank-pc-networks} reveal a clear geographical organization in the
estimated supports, although the regional labels enter neither the spectral
estimation nor the selection of the tuning parameters. This finding is
consistent with the strong geographical component of global bank equity
connectedness documented by \citet{Demirer2018} for the broader
96-bank panel. In a related frequency-domain analysis of the same
bank-volatility data, \citet{JimenezVaronKnight2026} found that the selected within-country bank
pairs generally exhibited stronger partial coherence, although some
cross-country pairs closely connected in the prespecified bank network
exhibited comparable values. These findings are qualitatively consistent
with both the pronounced within-region concentration and the cross-region
associations observed in our estimated networks. The differences among the three estimated networks further illustrate the
importance of examining bank-volatility dependence separately at different
frequencies, as emphasized by \citet{BarunikKrehlik2018}. Their analysis concerns directional connectedness
integrated over frequency bands and therefore provides motivation, rather
than a direct benchmark, for our pointwise partial-coherence results.

The pronounced regional concentration at the weekly frequency may
reflect banks' common exposure to regional market conditions over
shorter cyclical scales. The increased prevalence of cross-region
associations at the monthly and quarterly frequencies may be associated
with common global financial conditions operating over longer cycles.
These interpretations are tentative because the present analysis does
not separately identify the economic channels underlying the estimated
associations.

Section S7 of the Supplementary Material \citep{Supplement2026}
presents the partial coherence heatmaps and networks obtained using
CGLASSO-sc2 \citep{Deb2024}. Both methods use the same estimate of the
spectral density matrix at each target frequency and the same EBIC
setting. At the weekly frequency, the SGLASSO network shows clusters
of banks in the Americas, Europe, Asia, and Australia, providing a
more detailed view of regional conditional associations than the
CGLASSO-sc2 network. At the quarterly frequency, the two networks
exhibit similar geographical patterns.

\section{Conclusion}\label{sec:conclusion}
This paper develops SGLASSO for estimating sparse spectral precision
matrices in high-dimensional stationary time series. The theoretical
analysis establishes nonasymptotic error bounds for spectral density and
precision matrix estimation uniformly over the continuous frequency domain
and, under an additional minimum-signal condition, simultaneous exact
support recovery over any prescribed finite set of frequencies. In the
simulations, SGLASSO remained competitive in relative Frobenius error across
all designs and generally achieved higher AUROC values than CGLASSO, while
the F1 performance of the BIC-selected estimates varied across
data-generating processes. The estimated bank-volatility networks exhibited
frequency-dependent geographical organization: within-region connections
were most concentrated at the weekly frequency, whereas cross-region
connections were more prevalent at the monthly and quarterly frequencies.

Several questions remain for future research. The current procedure solves
the regularized inverse problem frequency by frequency, although the
first-stage estimator smooths the periodogram across neighboring
frequencies. It would be useful to develop theoretically justified,
data-driven bandwidth selection methods specifically targeted at spectral
precision matrix estimation. Joint regularization across neighboring
frequencies could also be investigated when the spectral precision matrix
varies smoothly or its support persists over local frequency bands. The
present theory focuses on estimation error and support recovery over finite
sets of frequencies rather than statistical inference. Developing entry-wise
confidence intervals and simultaneous tests for spectral precision entries
or partial coherence is therefore an important direction for future work.

\begin{acks}
AI tools were used to assist with grammar checking and proofreading. The authors take full responsibility for the content of the manuscript.
\end{acks}

\begin{funding}
This work is supported by the National Natural Science Foundation of China (Grant No.72273132), the Zhejiang Provincial Natural Science Foundation of China under Grant No. LMS26A010016, the National Key R\&D Program of China (No.2024YFA1013501), the Summit Advancement Disciplines of Zhejiang Province (Zhejiang Gongshang University - Statistics), and the Collaborative Innovation Center of Statistical Data Engineering Technology \& Application.
\end{funding}

\begin{supplement}
\stitle{Supplement to ``Sparse Spectral Precision Matrix Estimation for High-Dimensional Time Series''}
\sdescription{The supplement contains proofs of all theoretical results
presented in the main paper.}
\end{supplement}

\bibliographystyle{imsart-nameyear}
\bibliography{ref}

\end{document}